\documentstyle[aps,preprint,psfig]{revtex}

\newcommand{\NOO}{\mbox{NO$_2\,$}}
\newcommand{\grad}{\mbox{$^\circ$}}
\newcommand{\wavn}{\mbox{cm$^{-1}\,$}}
\begin{document}
\title{All the adiabatic bound states of \NOO (J=0)}
\author{R. F. Salzgeber$^\dagger$, V. Mandelshtam$^\star$, Ch. Schlier\footnote{Author to whom 
correspondence should be addressed}
$^\dagger$, H. S. Taylor$^\ddagger$}
\address{$^\dagger$Fakult\"at f\"ur Physik, Universit\"at Freiburg,
Hermann-Herder Stra{\ss}e 3, D-70104 Freiburg, Germany}
\address{$^\ddagger$University of Southern California, Los Angeles, CA 90089, USA}
\address{$^*$ Chemistry Department, University of California, Irvine, CA 92697, USA}
\date{received:$\phantom{01.01.0001}$}
\maketitle
\begin{abstract}
We calculated all 2967 even and odd bound states of the adiabatic ground state
of \NOO, using a modification of the {\it ab initio} potential energy surface
of Leonardi et al. [J. Chem. Phys. {\bf 105}, 9051 (1996)].
The calculation was performed by harmonic inversion of the Chebyshev
correlation function generated by a DVR Hamiltonian in Radau
coordinates.  The relative error for the computed eigenenergies (measured from
the potential minimum), is $10^{-4}$ or better, corresponding to an absolute 
error of less than about 2.5 \wavn.
Near the dissociation threshold the average density of states is about 
0.2/\wavn.
Statistical analysis of the states shows some interesting structure of the
rigidity parameter $\Delta_3$ as a function of energy.
\end{abstract}
Pacs 34.10.+x, 34.30.+h, 05.45.+b
\newpage
\section{Introduction}
The complexity of the spectrum of \NOO has always been a challenge, both to
experimentalists and theorists. Well-resolved spectra were first measured only
after the advent of the laser, and the introduction of
cooled molecular beams allowed the measurement of {\it complete} spectra in
certain energy ranges (still not up to the dissociation limit)
\cite{smalley:75,persch:88,georges:95a,georges:95b}. One reason for the
spectral congestion is the conical intersection (found by Gillespie {\it et
al.} in 1975 \cite{gillespie:75}) mixing the $\tilde{X}^2A_1$ electronic
ground state surface at an energy of only 1.289 eV with the first excited
$\tilde{A}^2B_2$ surface. 
The dissociation energy V$_{\rm diss}$ of $\tilde{X}^2A_1$ is 3.226 eV.
In the experiment, this leads to disturbed
progressions and increased state density, while the theorist is forced to
include the vibronic coupling in the calculation of more than the lowest few
vibrational states of the molecule \cite{haller:85,leitner:96}.

\noindent
There are only a few molecules for which anything like a full list of
vibrational states (even for J = 0) has been calculated. Examples are the
pioneering work of Tennyson et al.\cite{henderson:93} on H$_3^+$, and the more
recent calculations on HO$_2$\cite{ho2}, HCN\cite{bowman:97},
H$_3^+$\cite{h3+} and \NOO\cite{leonardi:96}. The reason for this situation is,
of course, the immense computational expense of such calculations, which one 
generally applies only to potential energy surfaces (PES) which are quite well 
known. These, however, are still scarce.

\noindent
In this paper we present first results of an effort to compute bound and
resonance states of \NOO. Two developments were of great help to us: first the
calculation of good PESs for the two lowest electronic states of \NOO including
their coupling by the Siena-Wuppertal group
\cite{leonardi:96}; second, the advent of
the filter diagonalization method
\cite{neuhauser:90,wall:95,grozdanov:95,mandelshtam:97a,mandelshtam:97b}
for solving large eigenvalue problems. The first step of this endavour, which
we report here, is the calculation of all ($\sim$3000) adiabatic bound states
of the ground PES. This in itself is non-trivial, since near the dissociation
limit the average state density is $\sim$0.2 states per \wavn for each of the 
two symmetry species. The computation of
non-adiabatic energy eigenvalues, which includes the surface coupling, is under
way\cite{salzgeber:98}, and the method itself allows the introduction of imaginary absorption
potentials, and, therefore, the extension of the calculations to resonances.
(This has been tested for H$_3^+$\cite{mandelshtam:97b})

\noindent
Another benefit of  this project is the possibility to perform a thorough
statistical analysis over the whole energy range of a molecular system.
This has so far been almost impossible for real molecules, because of missing
data in the experiments, or an insufficient number of eigenstates in the
calculation, both leading to large statistical
errors\cite{georges:95a,georges:95b,haller:85,leitner:96}.

\noindent
In the following we discuss first some properties and necessary modifications
to the PES of ref.\cite{leonardi:96}, then the computation, and, finally, some
results on the statistics of the computed levels.
\section{The potential energy surfaces}

Since the two lowest adiabatic PESs of \NOO have a conical intersection
at an energy of about one third of the dissociation energy, they must both be
considered in any calculation of more than the lowest few vibrational states
\cite{haller:85}. However, the intersection makes their direct interpolation
impractical. Instead, one interpolates the diabatic surfaces V$_{11}$ and
V$_{22}$, and, in addition, their coupling V$_{12}$. From such fits one
computes the energies of the two adiabatic surfaces V$_X$ (ground state) and
V$_A$ (first excited state) whenever they are needed from
\begin{eqnarray}
V^{\rm adiab}_{X/A} = {1\over 2}(V_{11} + V_{22})
\mp \sqrt{{1\over 4}(V_{11} -V_{22})^2 + V_{12}^2}
\end{eqnarray}
The surfaces used in our calculations have been derived in several steps from
nearly 1000 {\it ab initio} points computed in ref.\cite{hirsch:91}. These
points have been diabatized \cite{hirsch:90,hirsch:91} by a unitary transform,
demanding that certain dipole matrix elements be small. All computed points
are in the region (here we use bond coordinates) $2.08 a_0\le r_1, r_2\le2.50
a_0, 70\grad \le \beta\le 180\grad$, called region D in ref.\cite{hirsch:90}.
To produce surfaces useful for the calculation of vibrational levels, they had
to be fitted to an analytical form, furthermore they had to be extra\-polated 
to regions outside of region D, i.e. to shorter and larger interatomic 
distances, and to small angles. This has been done in several iterations by the
Siena and Wuppertal groups \cite{hirsch:91,leonardi:94,leonardi:96}, who also 
utilized empirical data
from the measured \NOO spectrum (see ref.[1] in \cite{leonardi:96}) to improve
the final fits of ref.\cite{leonardi:96}. It is these fits, called here
V$_{ik}^{\rm LP}$ (for Leonardi-Petrongolo) from which we started.

\noindent
While the authors of ref.\cite{leonardi:96} had no problems to compute
vibrational energies with their fits, early test calculations with our method
showed that the short- and long-range extrapo\-lations in V$_{ik}^{\rm LP}$ 
have (unphysical) kinks, which cannot be tolerated when a discrete variable
representation (DVR) is used. Therefore, we modified the surfaces by smoothing
out the kinks as described below. Preliminary calculations done by C.
Petrongolo \cite{privcom} with this surface using the code from ref.
\cite{leonardi:96}, and our nonadiabatic computation\cite{salzgeber:98} 
show an equally good agreement with experimental data at
energies up to 1.22 eV. Additionally, our surface modification improves the 
computed state density up to 2.3 eV, which had been too high compared with the
experimental one in ref.\cite{leonardi:96}.

\noindent
In detail,the following modifications were applied (energies are hartree units 
= 27.211396 eV, distances bohr units $a_0$ = 0.529177$\cdot10^{-10}$m):\\
{\bf (a)} The short-range correction (for $r_i < 2.08$) of the lower
diabatic surface $V^{\rm LP}_{11}$ was changed to (shown only for $r_1$, but
similar for $r_2$):
\begin{eqnarray}
V_{11}(r_1,r_2,\beta) &=& V_{11}^{LP}(r_1,r_2,\beta)(1-Q(a,b r_1))\\
   &+& (6.00(r_1-2.08)^2-V_{11}^{LP}(2.08,r_2,\beta)) Q(a,b r_1)\nonumber \\
   & & V_{11}^{LP} \in D, \nonumber
\end{eqnarray}
here the function $Q(a,br)=1-P(a,br)$ is the incomplete gamma function cf.
ref. \cite{recipes}) and the parameters a and b are listed in Table
\ref{switchdata}.\\
{\bf (b)} The short-range correction to V$_{22}$ was dropped, and V$_{22}(r_1,
r_2,\beta)$ set to a finite high value V$_{22}(1.50,r_2,\beta)$ for $r_1\le
1.50$, and similarly for $r_2\le 1.50$.\\
{\bf (c)} The small angle ($\beta < 70\grad$) corrections to the potentials
V$_{11}$ and V$_{22}$ were dropped completely.\\
{\bf (d)} For small and large angles ($\beta < 70\grad$ and $\beta >
131\grad$) the coupling potential was smoothly set to zero:
\begin{eqnarray}
V_{12}(\beta) &=& V_{12}^{LP}(\beta) (1-Q(a, b(\beta-71\grad)))
\mbox{\rm for}\; \beta < 71\grad \\
V_{12}(\beta) &=& V_{12}^{LP}(\beta) (1-Q(a, b(\beta-131\grad)))
\mbox{\rm for}\; \beta > 131\grad.
\end{eqnarray}
The values used for the parameters a and b are listed in Table
\ref{switchdata}.\\

{\bf (e)} The long-range correction of the diagonal potentials V$_{ii}$, which
are necessary to produce the  correct dissociation limits, where modified by
changing the switching function, now reading (for $r_1 > r_{\rm max} = 3.00$)
\begin{eqnarray}
V_{ii}(r_1,r_2) &=& (1-Q(1,b(r_1-r_{\rm max})) V_{ii}^{\rm LP} + \nonumber \\
     && Q(1,b(r_1-r_{\rm max})(V_{\rm diss} + V_{\rm NO}(r_2)).
\end{eqnarray}
and similar for $r_2$. The parameters a and b are again listed in table
\ref{switchdata}.

\section{Computation}
For computing bound states of \NOO  we use the low-storage version of the
filter-diagonalization method introduced recently by Mandelshtam and Taylor
\cite{mandelshtam:97a}. This method is conceptually based on the
filter-diagonalization procedure of Wall and Neuhauser \cite{wall:95}
which extracts the system eigenenergies by harmonic inversion of a time
correlation function C(t). The method of ref.~\cite{mandelshtam:97a} is
designed for a direct harmonic inversion of the Chebyshev correlation function
\cite{tal-ezer:84}
\begin{equation}
c_n =\langle\xi_0|T_n(\hat{H})|\xi_0\rangle\sim\sum_k d_k \cos n\omega_k\ ,
\end{equation}
for the eigenenergies $E_k=\cos\omega_k$ and amplitudes $d_k$.
The computation of the $c_n$ sequence is
done to  essentially the machine precision using a very inexpensive iterative
numerical scheme,
\begin{equation}\label{eq:cheb}
\xi_1=\hat H\xi_0,\;\;\;\xi_{n+1}=2\hat H\xi_n-\xi_{n-1}\ ,
\end{equation}
with $c_n$ being generated using
$c_{2n}=2\langle\xi_n|\xi_n\rangle-c_0,\ \  c_{2n-1}=2\langle\xi_{n-1}|\xi_n
\rangle-c_1$.
This requires to store only a few vectors at a time, assuming that the
matrix-vector multiplication is implemented without explicit storage of the
Hamiltonian matrix. The spectral analysis part (i.e., the harmonic inversion
of $c_n$ by the filter-diagonalization) is carried out independently and
computationally very
efficiently after the sequence $c_n$ is computed. All these features assure the
very high performance of the overall numerical procedure.

\noindent
For the Hamiltonian matrix representation we choose Radau coordinates ($R_1, 
R_2, \Theta$) in
which all mixed derivatives in the kinetic energy vanish \cite{sutcliffe:91},
to account for the C$_{2v}$ symmetry of the \NOO molecule.
The fast application of the Hamiltonian matrix to a vector is achieved by 
implementing a potential adapted DVR, which we set up in a way which allowed
us to identify a single parameter, the potential cutoff energy $V_{cut}$,
on which all other grid parameters depend. The bases for the radial parts of 
the Hamiltonian are sinc-DVRs\cite{colbert:92}, characterized by their spatial 
extensions $R_{i\,min}$ to $R_{i\,max}$ (i = 1, 2), and sizes $n_1  = n_2$, 
which are diagonalized and then truncated by an energy cutoff to $n_{1b} = 
n_{2b}$. For the angular part we use a Legendre-DVR\cite{light:85} 
characterized by its size $n_3$. The angular extension is always 0\grad to 
180\grad.

\noindent
Convergence studies in which all grid parameters are varied independently are 
prohibited in this study by the computing times involved. Instead, these 
parameters were fixed in 1D calculations, but in a consistent way: We first 
set an accuracy goal, in this case a relative accuracy of $10^{-4}$ for states 
just below the true \NOO dissociation energy of 3.226 eV. With this desired 
accuracy we adjust (in 1D calculations at the equilibrium angle $\Theta_0$)
the spatial extensions $R_{i\,min}$ to $R_{i\,max}$ (i = 1, 2) of the radial 
grids, while the sizes $n_i$ are kept very large. Then the sizes of the 
spatial grids, $n_1 = n_2$, are adjusted to yield this same desired accuracy, 
similarly for the angular grid (at $R_1 = R_2 = R_0$), whose extension, 0\grad 
to 180\grad, is kept. 
The equidistant grid spacings of the 1D-DVR points $\Delta R, \Delta\Theta$, 
on the other hand, determine a maximum kinetic energy, 
\begin{eqnarray}
T^{\rm radial}_{\rm max}={\pi^2\over2\mu(\Delta R_0)^2}\,\,{\rm and}\\
T^{\rm angular}_{max}\simeq
{\pi^2\over2\mu}\left({1\over R_1^2}+{1\over R_2^2}\right)
{1\over\Delta\Theta^2}.\label{Tang}
\end{eqnarray}
(We used $R_1 = R_0$ and $R_2 = \infty$ as an estimate for the moment of 
inertia in eq. \ref{Tang}, which seems reasonable for an almost dissociating 
wavepacket.)

\noindent
The larger one of these kinetic energies, which in our example ($10^{-4}$ 
relative accuracy at 3.226 eV) are 4.5 eV for the radial grids and 8.0 eV for 
the angular grid, is then used to fix the potential cutoff $V_{cut}$. For 
consistency, the radial grids are increased in size such that T$_{\rm max}
^{\rm radial} = T_{\rm max}^{\rm angular}$. Then all DVR-points for which 
V($R_1,R_2,\Theta) > V_{cut}$ are deleted. As a final result for the accuracy 
goal mentioned above we get a primitive grid of $n_1 = n_2$ = 127, $R_{1\,min}
= R_{1\,max} = 1.5 a_0, R_{1\,max} = R_{2\,max} = 5.5 a_0, n_3 = 275$, a 1D 
truncation of the radial grids to $n_{1b} = n_{2b}$ = 100, and a final 
truncated grid of 1 112 707 points. 

\noindent
A further parameter of the method is the number of iterations of the filter 
diagonalization, eq. \ref{eq:cheb}, which is one half of the number of 
Chebyshev coefficients $c_n$. In addition, for the final stage of the 
calculation the window parameters $E_{min}, E_{max}$, and the number of window 
basis functions (described in ref.\cite{mandelshtam:95}) have to be fixed. 
This was done such that the extracted eigenenergies were converged to 6 
figures with respect to the sequence of coefficients $c_n$, i.e. two orders of 
magnitude better than our overall accuracy goal. This led to a number of 
300~000 coefficients, whose calculation for the 8 eV set of one symmetry 
needed about 30 days of CPU-time on an IBM RS6000/59H.

\noindent
To ensure that our overall accuracy is in the exponential convergence region, 
and not in some --- misleading --- small local convergence minimum 
\cite{seidemann:92}, we ran four different grids consistent with energy 
cutoffs of 4, 6, 8 and 8.1 eV. The last one has been used in comparison to the 
last but one to observe numerical noise, and to have an independent set of 
eigenvalues ensuring that we do not miss any levels. All inconsistencies 
between the two largest sets have been cleared.

\noindent
The errors of the computed eigenvalues can be approximately taken as the noisy 
difference between the calculation with $V_{cut}$ = 8.0 eV and the one with 
$V_{cut}$ = 8.1 eV. As Fig. \ref{converg} shows, it increases with energy 
above about 2 eV, and above 3 eV has a plateau of about $10^{-4}$ eV. This 
plateau, which is also shown by the differences of the smaller calculations 
(with $V_{cut}$ = 4.0 and 6.0 eV, respectively), is caused by the common 
limited spatial extension of the radial basis sets, which do not cover the 
tails of the high energy wavefunctions exceeding $R_{i\,max}$. Due to a 
smaller density of DVR points in these bases, their plateau starts earlier 
than that of $V_{cut}$ = 8.0 eV. Apart from that, the errors of the smaller 
calculations lie in the expected order. Results for the odd states are similar.
The fact, that all errors lie below the desired precision, proves the 
convergence of the 3D computation.

\section{Statistical analysis of the energy levels}
\noindent
Our result consists of the energies of 1606 even and 1361 odd levels of \NOO 
\cite{PAPS}. (Note that the symmetries of this adiabatic calculation have no 
physical meaning for the true non-adiabatic molecule with its dominant conical 
intersection). At energies below 10~000\wavn they agree with the published 
list of experimental values\cite{delon:91a} to within an average error of 14
\wavn. Since in that energy region the potential fit has been adjusted using 
the experimental data, this shows the correctness of our calculation in this 
energy range. Similar agreement exists between our computed data and those 
computed in ref.\cite{leonardi:96}.

\noindent
At much higher energies a comparison with experimental data is questionable 
due to the inaccuracies of the extrapolated PESs, which we had to use, and the
adiabatic approximation. However, statistical analyses are meaningful. So, we 
have performed separate statistical analyses of the 1606 even and 1361 odd 
levels. One expects a transition from regular to irregular dynamics with 
rising energy not later than the conical intersection at 1.289 eV. Since only 
about 75 even and 50 odd states lie below
this energy, nearest neighbor histograms trying to show this effect, suffer
from insufficient data. Nevertheless, in Fig. \ref{nearneigh} the nearest
neighbor distributions of the even states are shown for four different energy
windows covering all parts of the spectrum. All distributions imply chaotic 
dynamics by their Wigner type shape, but only the two high energy distributions
contain enough data to be really trustworthy. One may, perhaps, wonder why the 
adiabatic, uncoupled, ground state PES of \NOO shows "chaos" at such a low 
energy. However, one should not forget that (a) any anharmonic PES shows chaos 
gradually increasing with energy, and (b) that the low-lying conical 
intersection in \NOO {\it indirectly} distorts the shape of the {\it adiabatic}
surfaces very much compared with, e.g., a triple Morse potential. The 
contribution of the Born-Oppenheimer breakdown, i.e. the {\it direct} coupling 
between the surfaces as such, does not change the level statistics too 
much\cite{salzgeber:98}.

\noindent
The rigidity of the spectrum, $\Delta_3(L)$ seems to be more robust. It is
defined as the local mean square deviation of the level staircase N(E) from the
best fitting straight line over an energy range $[E_i,E_{i+L}]$ corresponding
to L mean level spacings, namely
\begin{eqnarray}
\Delta_3(L) = \stackrel{\rm min}{\scriptscriptstyle(a,b)}
\left\{ {1\over L}
\int_{E_i}^{E_{i+L}} [N(E) -aE -b]^2 dE \right\}.
\end{eqnarray}

\noindent
We have calculated it for lengths L between 2 and 300 levels for the
whole spectrum, and for several fractions of it. We have also used various
unfolding procedures \cite{haller:83} (and no unfolding at all, which has
no effect at energies above 2.3 eV). We find no energy range in which
$\Delta_3(L)$ is near L/15, the expectation value for random level positions,
which one generally correlates with regular classical dynamics. Instead,
taking the usual values of L from 2 to 30, we find $\Delta_3(L)$ slightly
above the expectation for a Gaussian orthogonal ensemble (GOE), but
never higher than twice that value. Interestingly, when we looked for the
fluctuations at a much wider scale, taking L between 100 and 300, we found an
energy dependent behavior, which, in addition, differed for the even and odd
states.

An example is shown in Fig. \ref{delta3}. For L = 100, the expected values for
random and GOE sequences are 6.67 and 0.46, respectively. Our results lie
between 0.5 and 2.5 in different parts of the spectrum. They are robust
against a variation of L between 100 and 300, and against random shifts of the
energy levels of up to 0.5\%, simulating errors in the diagonalization
procedure. To demonstrate that Fig. \ref{delta3} indeed shows the developement
of the dynamical structure of the states with energy, we show the more
conventional way of computing the averaged statistics $\langle\Delta_3(L)
\rangle$ in Fig. \ref{delta3b} for two energy ranges: $0 \le E \le 1.3$eV
below the conical
intersection, and  $2.5 \le E \le 2.8$eV, where the difference between
the symmetries is most prominent. The averaged statistics confirms the data of
Fig. \ref{delta3} as representative. Therefore, we are convinced that these 
changes of the rigidity of the spectrum with energy are a property of the 
spectrum of \NOO, or at least of this model of it.

\noindent
The saturation of the curves in Fig. \ref{delta3b}b was explained by
Berry\cite{berry:85} in a semiclassical treatment of an integrable and a
totally chaotic system. According to this paper, the value of $L_{\rm max}$ at
which saturation takes place is associated with the period of the shortest 
classical orbit of the system. At this moment, it is unclear to us, why 
$L_{\rm max}$ is different for even and odd symmetries.

\noindent
In conclusion, we have shown that the calculation of all J=0-states of a
non-hydride second row molecule having a few thousand states is feasible, and
we foresee that the same holds in the presence of a conical intersection, and
for low lying resonances. No supercomputer is needed if one is willing to
pay the price of some patience.

\acknowledgments This work was supported by the Deutsche Forschungsgemeinschaft
through the SFB 276 ``Korrelierte Dynamik hochangeregter atomarer und
molekularer Systeme'' and by the Max Planck prize granted to H. S. Taylor
and J. Briggs. We are grateful to C. Petrongolo for the permission to use the
potential fit of ref.\cite{leonardi:96}, and for some test calculations. RFS
wishes to thank V. Mandelshtam, H. S. Taylor and C. Petrongolo for their
hospitality and helpful discussions. We also want to express our thanks
to U. Manthe, J. Tennyson and J. Schryber for helpful remarks.
HST wishes to acknowledge support from DOE (contract number 53-4815-2480).

\newpage

\newpage

\begin{figure}
\caption{The reciprocal of the median (taken from a window of 100 eigenvalues) 
of the density of even states $\rho(E)$, and the median of the absolute errors 
of the energy eigenvalues of the various computations with respect to the 
largest calculation (with $V_{cut} = 8.1$ eV) {\it vs.} energy.}
\label{converg}
\end{figure}

\begin{figure}
\caption{Nearest neighbor distributions of the even states for four
different energy ranges: (a) below the conical intersection, (b) about the
conical intersection, (c) slightly, and (d) high above the conical
intersection. The best fits to Poisson and Wigner distributions are
also shown.}
\label{nearneigh}
\end{figure}

\begin{figure}
\caption{$\Delta_3$ statistics for a window of 100 even (solid line) and odd
(dashed line) levels shifted
in steps of 10 over the energy range. The $\Delta_3$ value is assigned to the
energy in the middle of the window. The horizontal line is the value of
the corresponding $\Delta_3$ for a Gaussian orthogonal ensemble (GOE).}
\label{delta3}
\end{figure}

\begin{figure}
\caption{The averaged $\Delta_3$ statistics and its rms error for a
window of L states moved between (a) 0 and 1.3 eV, and (b) 2.5 and 2.8 eV.
The expectation values for Poisson statistics (P) and the Gaussian orthogonal
ensemble (GOE) are also shown.}
\label{delta3b}
\end{figure}

\newpage
\begin{table}
\caption{Values of the parameters a and b of the incomplete Gamma
function used to switch between potential parts.}
\label{switchdata}
\begin{tabular}{lcc}
correction           &   a   &  b [${a_0}^{-1}$]  \\ \tableline
short-range $V_{11}$ & 1.1   & -30.0   \\
long-range $V_{11}$  & 3.5   & 6.34902 \\
long-range $V_{22}$  & 2.0   & 2.09979 \\
angle $V_{12}$       & 2.0   & 20.0
\end{tabular}
\end{table}

\newpage
\centering
\pagestyle{empty}
\psfig{figure=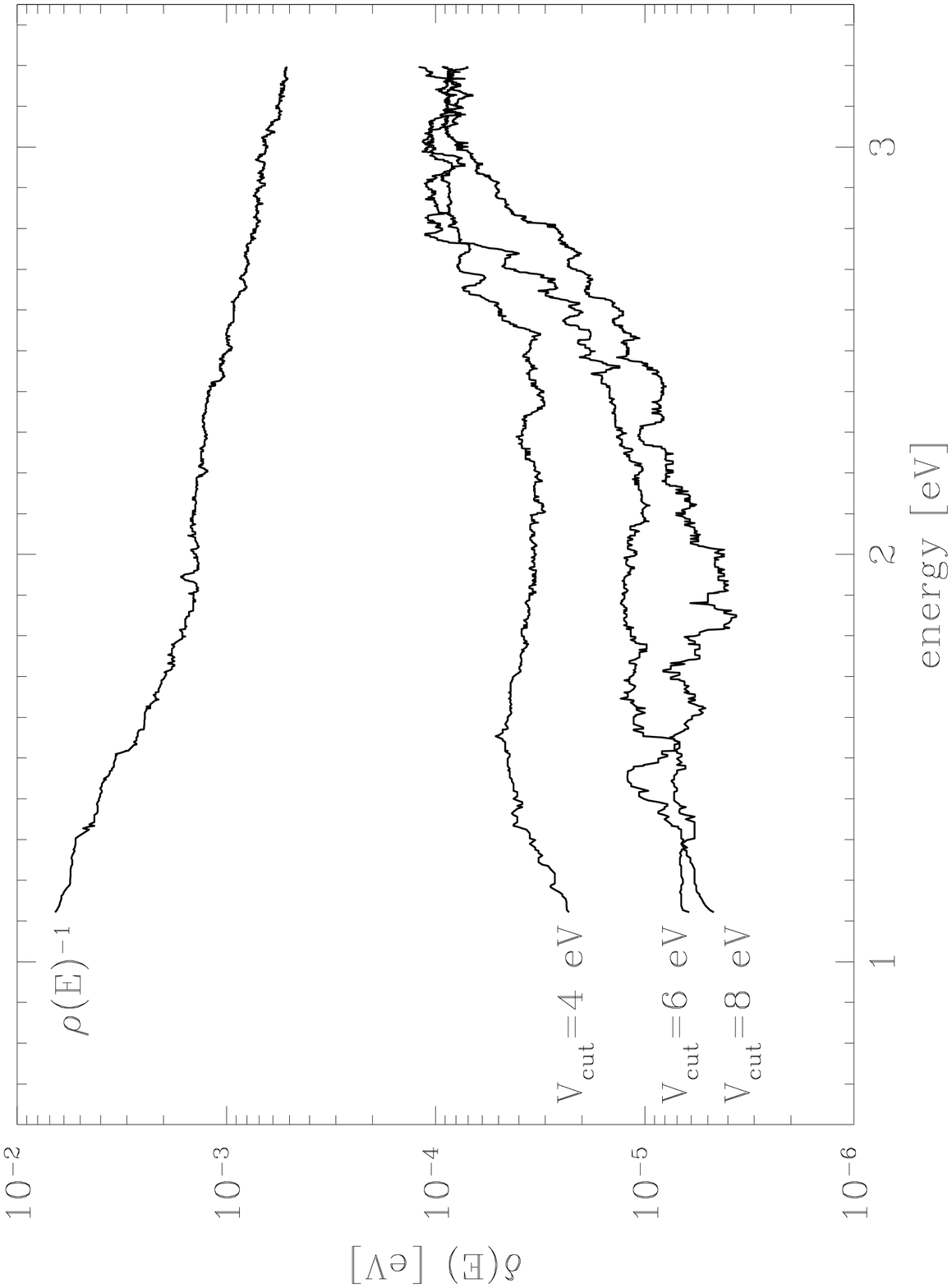,width=16cm}
\newpage
\centering
\psfig{figure=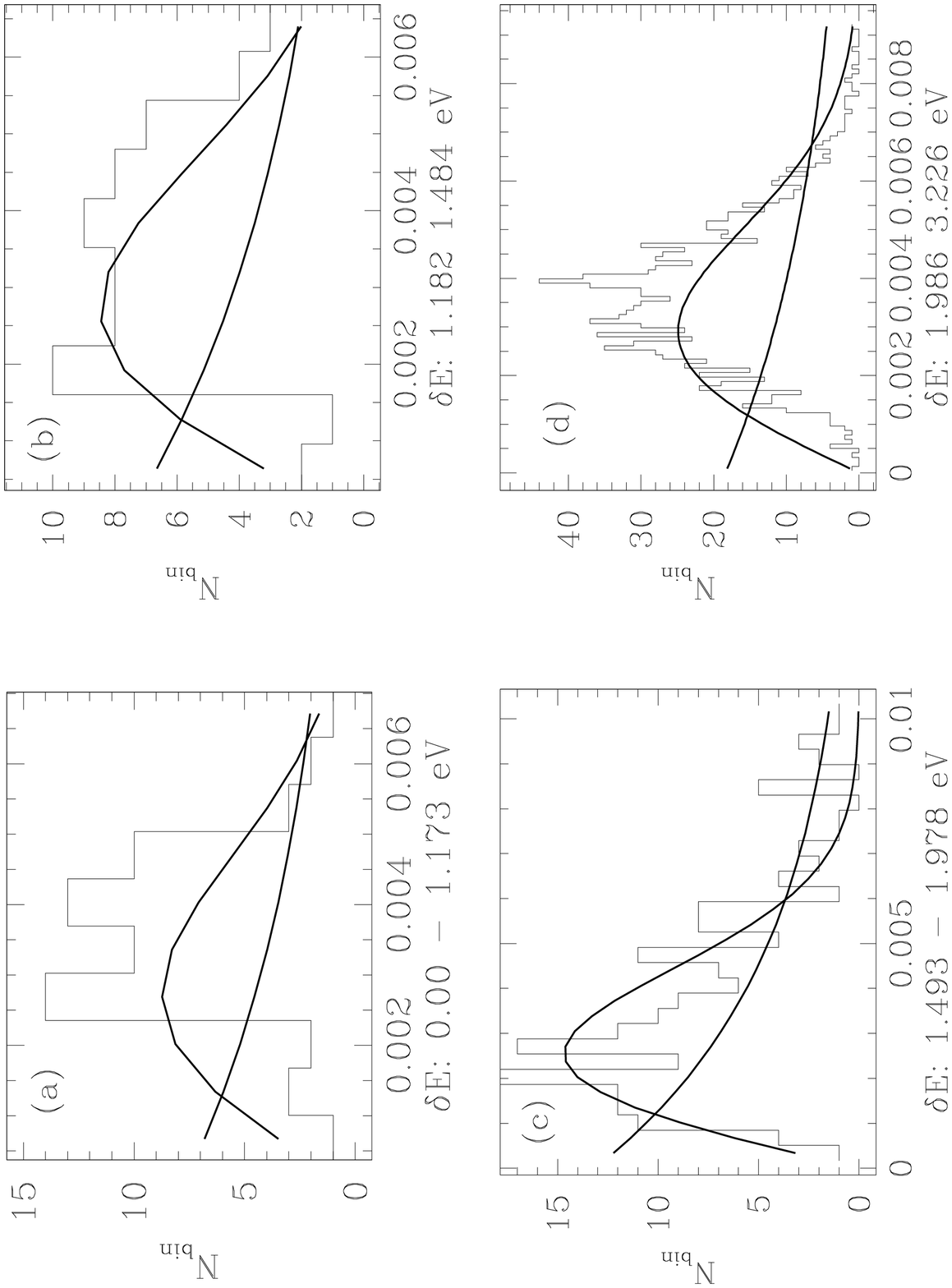,width=16cm}
\newpage
\centering
\psfig{figure=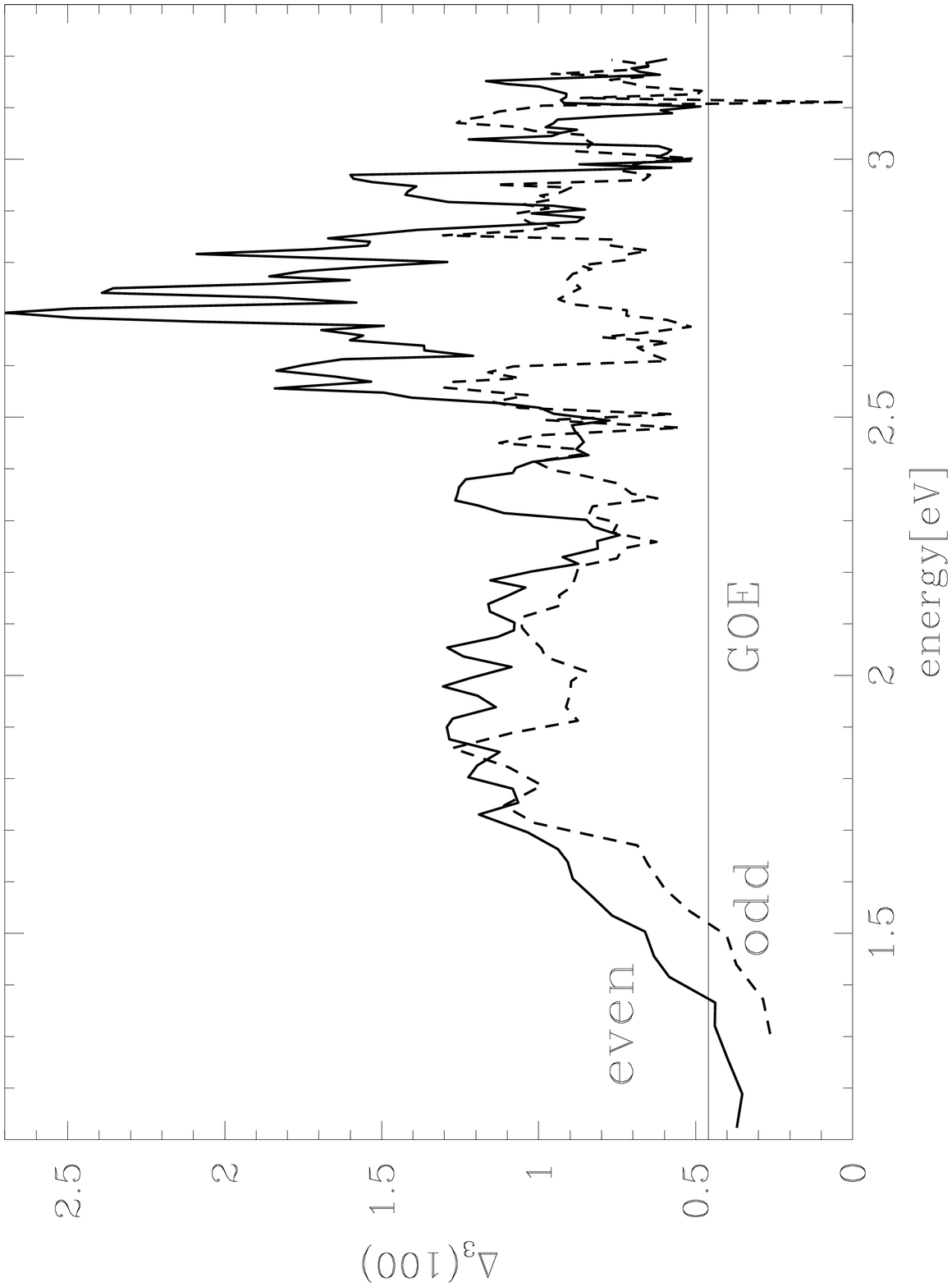,width=16cm}
\newpage
\centering
\psfig{figure=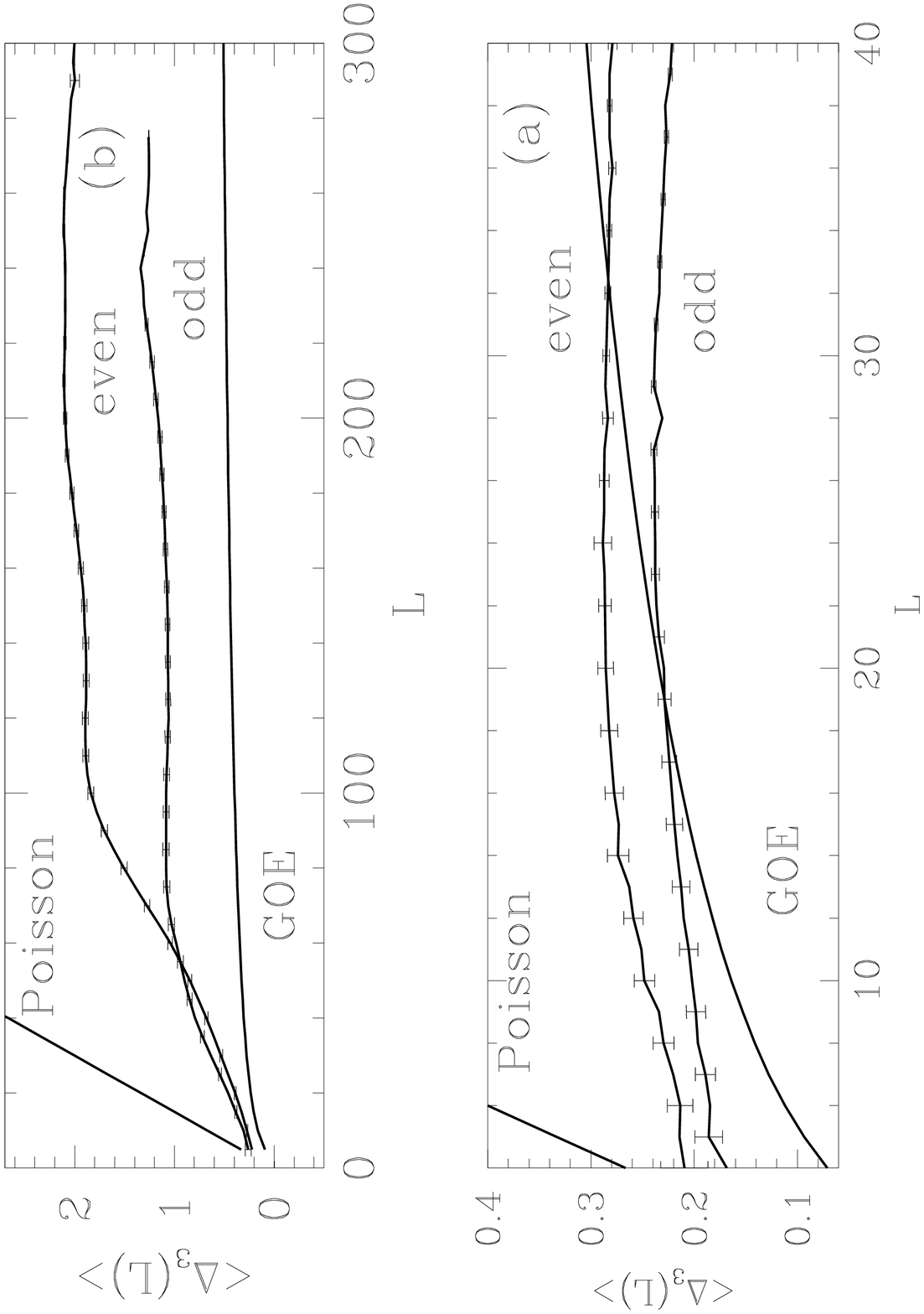,width=16cm}

\end{document}